# CONTROLLED STRATIFICATION FOR QUANTILE ESTIMATION

By Claire Cannamela, Josselin Garnier and Bertrand Iooss

*Commissariat à l'Energie Atomique, Université Paris VII and Commissariat à l'Energie Atomique*

In this paper we propose and discuss variance reduction techniques for the estimation of quantiles of the output of a complex model with random input parameters. These techniques are based on the use of a reduced model, such as a metamodel or a response surface. The reduced model can be used as a control variate; or a rejection method can be implemented to sample the realizations of the input parameters in prescribed relevant strata; or the reduced model can be used to determine a good biased distribution of the input parameters for the implementation of an importance sampling strategy. The different strategies are analyzed and the asymptotic variances are computed, which shows the benefit of an adaptive controlled stratification method. This method is finally applied to a real example (computation of the peak cladding temperature during a large-break loss of coolant accident in a nuclear reactor).

**1. Introduction.** Quantile estimation is of fundamental importance in statistics as well as in practical applications [Law and Kelton (1991)]. A main challenge in quantile estimation is that the number of observations can be limited. This is the case when the observations correspond to the output of expensive numerical simulations. Variance reduction techniques specifically designed for this problem have been proposed and implemented. These techniques usually involve importance sampling [Glynn (1996)], correlation-induction [Avramidis and Wilson (1998)] and control variates [Hsu and Nelson (1987), Hesterberg and Nelson (1998)]. In this paper we focus our attention on variance reduction techniques based on the use of an auxiliary variable. For the problem of the quantile estimation of a computer code output, the auxiliary variable is the output of a reduced model, which is coarse but cheap from the computational time point of view. We will show how to use it to find convenient parameters for the stratified and importance sampling techniques [Rubinstein (1981)].

Quantiles form a class of performance measures. Quantile estimation for a real-valued random variable (r.v.) $Y$ aims at determining the level $y$ such that the likelihood that $Y$ takes a value lower than $y$ is some prescribed value. We assume that $Y$ has an absolutely continuous cumulative distribution function (cdf) $F(y) = \mathbb{P}(Y \leq y)$ and a continuously differentiable probability density (pdf) $p(y)$. We look for an estimation of the $\alpha$-quantile $y_\alpha$ defined by $F(y_\alpha) = \alpha$.

In this paper we assume that $Y = f(X)$ is the real-valued output of a computer code $f$ that is CPU time expensive and whose input parameters are random and modeled by the random vector $X \in \mathbb{R}^d$ with known distribution. With the advance of computing technology and numerical methods, the design, modeling and analysis of computer code experiments have been the subject of intense research during the last two decades [Sacks et al. (1989), Fang, Li and Sudjianto (2006)]. The problem of low or high quantile estimation (smaller than 10% and larger than 90%) can be resolved by classical sampling techniques such as the simple Monte Carlo or Latin hypercube techniques. These Monte Carlo methods can give imprecise quantile estimate (with large variance) if they are performed with a limited number of code runs, say, of the order of 200 [David (1981)]. An alternative approach is, to calculate a tolerance limit rather than a percentile by using Wilk's formula [Nutt and Wallis (2004)]. It provides with a low number of code runs, less than 200, say, a majoring value of the desired percentile with a given confidence level (e.g., 95%). But the variance of this tolerance limit is larger than that of the empirical estimate, for the same number of code runs.

Another well known approach for the uncertainty analysis of complicated computer models consists in replacing the complex computer code by a reduced model, called metamodel or response surface [Fang, Li and Sudjianto (2006)]. However, a low or high quantile estimate from a metamodel tends to be substantially different from the full computer model quantile because the metamodel is usually constructed by smoothing of the computer model output. Recently, some authors have taken advantage of one particular type of metamodel: the Gaussian process model [Sacks et al. (1989)], which gives not only a predictor (the mean) of a computer experiment but also a local indicator of prediction accuracy (the variance). In this context, Oakley (2004), Vazquez and Piera Martinez (2008) and Ranjan, Bingham and Michailidis (2008) have developed sequential procedures to choose design points and to construct an accurate Gaussian process metamodel, specially near the regions of interest, where the quantile lies. Rutherford (2006) proposes to use geostatistical conditional simulation techniques to obtain many realizations of the Gaussian process, which in turn can give a quantile estimate. However, all these techniques are based on the construction of a Gaussian process model which can be difficult, albeit possible [Jones, Schonlau and Welch (1998), Schonlau and Welch (2005), Marrel et al. (2008)], in the high-dimensional context ($d > 10$). Moreover, in industrial practice, a metamodel may already be available that comes from a previous study or from a simplified physical model. This is the situation we have in mind. We do not concentrate our effort on the construction of a more accurate metamodel, but on the use of a given reduced model.

In our work we deal with this situation in which a reduced model is available, in the form of a metamodel $f_r$, which is a coarse approximation of the computer model $f$. The quality of the metamodel may not be known; the metamodel may be a simplified version of the computer code (a one-dimensional version for a three-dimensional problem, a response surface determined during another study,...); its input variables may be a subset of the input variables for the computer code $f$. The full computer model $f$ is assumed to be very computationally expensive, while the evaluation of the metamodel $f_r$ and the generation of the input r.v. $X$ are assumed to be very fast (essentially free). Therefore, the focus of this paper is on how to exploit the metamodel to obtain better control variates, stratification or importance sampling than could be obtained without it. In the real example we address in Section 5 (computation of the peak cladding temperature of the nuclear fuel during a large-break loss of coolant accident in a nuclear reactor), the CPU time for each call of the function $f$ is 20 minutes, while the metamodel $f_r$ is a linear function and the input r.v. $X$ is a collection of $d = 53$ independent real-valued r.v. with normal and log-normal distributions. In this example, we also study the relevance of using more complex and powerful metamodels, such as the popular Gaussian process model.

The usual practice of quantile estimation is to construct an estimator of the cdf of $Y$ first, then to deduce an estimator of the $\alpha$-quantile of $Y$. In absence of the control variate, the standard method is the following one. The estimation is based on a $n$-sample $(Y_1, \ldots, Y_n)$, that is to say, a set of $n$ independent and identically distributed r.v. with the pdf $p(y)$ of $Y$. The empirical estimator (EE) of the cdf of $F$ is

$$\hat{F}_{\text{EE}}(y) = \frac{1}{n} \sum_{i=1}^{n} \mathbf{1}_{Y_i \leq y}, \tag{1}$$

which leads to the standard estimator of the $\alpha$-quantile

$$\hat{Y}_{\text{EE}}(\alpha) = \inf\{y, \hat{F}_{\text{EE}}(y) > \alpha\} = Y_{(\lceil \alpha n \rceil)}, \tag{2}$$

where $\lceil x \rceil$ is the integer ceiling of $x$ and $Y_{(k)}$ is the $k$th order statistics. Refined versions of this result based on interpolation and smoothing methods can be found in the literature [Dielman, Lowry and Pfaffenberger (1994)]. The empirical estimator $\hat{Y}_{\text{EE}}(\alpha)$ has a bias and a variance of order $1/n$

[David (1981)]. The empirical estimator is asymptotically normal,

$$\sqrt{n}(\hat{Y}_{\text{EE}}(\alpha) - y_\alpha) \stackrel{n \to \infty}{\longrightarrow} \mathcal{N}(0, \sigma_{\text{EE}}^2), \qquad \sigma_{\text{EE}}^2 = \frac{\alpha(1-\alpha)}{p^2(y_\alpha)}. \tag{3}$$

We remark that the reduced variance $\sigma_{\text{EE}}^2$ is usually larger when a larger quantile is estimated (the pdf at $y_\alpha$ is then very small).

The outline of the paper is as follows. First we describe quantile estimation by control variate in Section 2. Section 3 presents an original controlled stratification method. Then, a controlled importance sampling strategy is analyzed in Section 4. A real example is addressed in Section 5.

**2. Quantile estimation by Control Variate (CV).** In this section we present the well-known variance reduction techniques based on the use of $Z = f_r(X)$ as a control variate. The quantiles $z_\alpha$ of $Z$ are assumed to be known, as well as any expectation $\mathbb{E}[g(Z)]$ of a function of $Z$. We mean that these quantities can be computed analytically, or they can be estimated by standard Monte Carlo estimations with an arbitrary precision, since only the reduced model $f_r$ is involved.

2.1. *Estimation of the distribution function.* A Control Variate (CV) estimator of $F(y)$ with the real-valued control variate $Z$ has the form

$$\hat{F}_{\text{CV}}(y) = \hat{F}_{\text{EE}}(y) - C(\hat{g}_n - \mathbb{E}[g(Z)]), \tag{4}$$

where the function $g: \mathbb{R} \to \mathbb{R}$ has to be chosen by the user [Nelson (1990)] and $\hat{g}_n = \frac{1}{n} \sum_{i=1}^n g(Z_i)$. The optimal parameter $C$ is the correlation coefficient between $g(Z)$ and $\mathbf{1}_{Y \leq y}$ whose value is unknown in practice. Therefore, the estimated parameter $\hat{C}$ is used instead. It is defined as the slope estimator obtained from a least-squares regression of $\mathbf{1}_{Y_j \leq y}$ on $g(Z_i)$:

$$\hat{C} = \frac{\sum_{j=1}^n (\mathbf{1}_{Y_j \leq y} - \hat{F}_{\text{EE}}(y))(g(Z_j) - \hat{g}_n)}{\sum_{j=1}^n (g(Z_j) - \hat{g}_n)^2}.$$

As shown by Hesterberg (1993), the estimator $\hat{F}_{\text{CV}}(y)$ with the estimated parameter $\hat{C}$ can be rewritten as the weighted average

$$\hat{F}_{\text{CV}}(y) = \sum_{j=1}^n W_j \mathbf{1}_{Y_j \leq y} \tag{5}$$

with

$$W_j = \frac{1}{n} + \frac{(\hat{g}_n - \mathbb{E}[g(Z)])(\hat{g}_n - g(Z_j))}{\sum_{i=1}^n (g(Z_i) - \hat{g}_n)^2}.$$

Note that $\sum_{j=1}^{n} W_j = 1$. If $g(z) = \mathbf{1}_{z \leq z_\alpha}$, then $\mathbb{E}[g(Z)] = \alpha$, $\hat{g}_n = N_0/n$ with

$$(6) \qquad N_0 = \sum_{i=1}^{n} \mathbf{1}_{Z_i \leq z_\alpha} \quad \text{and} \quad W_j = \frac{\alpha}{N_0} \mathbf{1}_{Z_j \leq z_\alpha} + \frac{1-\alpha}{n-N_0} \mathbf{1}_{Z_j > z_\alpha}.$$

As shown by Davidson and MacKinnon (1992), the estimator (5) is equivalent to the maximum likelihood estimator for probabilities. By using standard results for the convergence of Monte Carlo estimators [Nelson (1990)], one finds

$$(7) \qquad \sqrt{n}(\hat{F}_{\mathrm{CV}}(y) - F(y)) \stackrel{n \to \infty}{\longrightarrow} \mathcal{N}(0, \sigma_{\mathrm{CV}}^2),$$
$$\sigma_{\mathrm{CV}}^2 = F(y)(1 - F(y))(1 - \rho_I^2),$$

where $\rho_I$ is the correlation coefficient between $\mathbf{1}_{Y \leq y}$ and $\mathbf{1}_{Z \leq z_\alpha}$:

$$(8) \qquad \rho_I = \frac{\mathbb{P}(Y \leq y, Z \leq z_\alpha) - \alpha F(y)}{\sqrt{F(y)(1 - F(y))}\sqrt{\alpha - \alpha^2}}.$$

This result can be compared to the corresponding central limit theorem in absence of control, which claims that the empirical estimator $\hat{F}_{\mathrm{EE}}$ defined by (1) is asymptotically normal,

$$\sqrt{n}(\hat{F}_{\mathrm{EE}}(y) - F(y)) \stackrel{n \to \infty}{\longrightarrow} \mathcal{N}(0, \sigma_{\mathrm{EE}}^2), \qquad \sigma_{\mathrm{EE}}^2 = F(y)(1 - F(y)).$$

Comparing with (7) reveals an asymptotic variance reduction of $1 - \rho_I^2$.

2.2. *Quantile estimation.* Our goal is now to estimate the $\alpha$-quantile of $Y$ by using the CV estimator of the cdf of $Y$. We consider the order statistics $(Y_{(1)}, \ldots, Y_{(n)})$ with the weights $W_{(i)}$ defined by (6) sorted according to the $Y_{(i)}$. Using the estimator (5) of the cdf of $Y$, the CV estimator of the $\alpha$-quantile is

$$(9) \qquad \hat{Y}_{\mathrm{CV}}(\alpha) = Y_{(K)}, \qquad K = \inf\left\{j, \sum_{i=1}^{j} W_{(i)} > \alpha\right\}.$$

Applying standard results for the variance reduction for Monte Carlo methods [David (1981)], one finds that this estimator is asymptotically normal with the reduced variance $\sigma_{\mathrm{CV}}^2$,

$$(10) \quad \sqrt{n}(\hat{Y}_{\mathrm{CV}}(\alpha) - y_\alpha) \stackrel{n \to \infty}{\longrightarrow} \mathcal{N}(0, \sigma_{\mathrm{CV}}^2), \qquad \sigma_{\mathrm{CV}}^2 = \frac{\alpha(1-\alpha)}{p^2(y_\alpha)}(1 - \rho_I^2),$$

where $p$ is the pdf of $Y$ and $\rho_I$ is the correlation coefficient between $\mathbf{1}_{Y \leq y_\alpha}$ and $\mathbf{1}_{Z \leq z_\alpha}$:

$$\rho_I = \frac{\mathbb{P}(Y \leq y_\alpha, Z \leq z_\alpha) - \alpha^2}{\alpha - \alpha^2}.$$

Comparing (10) with (3) reveals a variance reduction by the factor $1 - \rho_I^2$. As expected, the stronger $Y$ and $Z$ are correlated, the larger the variance reduction is. It is not easy to build an estimator of the reduced variance $\sigma_{\text{CV}}$, because this requires to estimate the pdf $p(y_\alpha)$. However, it is possible to build an estimator of the correlation coefficient $\rho_I$, which controls the variance reduction. This estimator is the empirical correlation coefficient $\hat{\rho}_I$ defined by

$$(11) \quad \hat{\rho}_I = \frac{\sum_{j=1}^n (\mathbf{1}_{Y_j \leq y} - \hat{F}_{\text{EE}}(y))(\mathbf{1}_{Z_j \leq z_\alpha} - \hat{G}_n(z_\alpha))}{\sqrt{\sum_{j=1}^n (\mathbf{1}_{Y_j \leq y} - \hat{F}_{\text{EE}}(y))^2} \sqrt{\sum_{j=1}^n (\mathbf{1}_{Z_j \leq z_\alpha} - \hat{G}_n(z_\alpha))^2}} \Big|_{y = \hat{Y}_{\text{CV}}(\alpha)}$$

with $\hat{G}_n(z_\alpha) = \frac{1}{n} \sum_{i=1}^n \mathbf{1}_{Z_i \leq z_\alpha}$.

2.3. *The optimal CV estimator.* In the previous section the control variate function is $g(z) = \mathbf{1}_{z \leq z_\alpha}$, which allows both an easy implementation and a substantial variance reduction. In general, the variance reduction obtained with the CV estimator (4) depends on the correlation coefficient between $\mathbf{1}_{Y \leq y}$ and the control $g(Z)$. The optimal control, which maximizes the correlation coefficient, is obtained with the function [Rao (1973)]

$$(12) \qquad g^*(z) = \mathbb{P}(Y \leq y | Z = z).$$

This function is usually unknown, otherwise it could be possible to compute analytically the cdf $F(y)$ by taking the expectation of $g^*(Z)$, and solve numerically the equation $F(y) = \alpha$ to get the quantile. However, this result gives the principle of refined CV methods using approximations of the optimal control function $g^*$. Continuous approximations have been proposed, that are however difficult to implement in practice [Hastie and Tibshirani (1990)]. Discrete approximations have been presented, which have been shown to be very efficient and easy to implement. We now describe the discrete method proposed by Hesterberg and Nelson (1998). Let us choose $m+1$ cutpoints $0 = \alpha_0 < \alpha_1 < \cdots < \alpha_m = 1$, and denote by $-\infty = z_{\alpha_0} < z_{\alpha_1} < \cdots < z_{\alpha_m} = \infty$ the corresponding quantiles of $Z$. The intervals $(z_{\alpha_{j-1}}, z_{\alpha_j}]$ will be used as strata to construct a stepwise constant approximation of the optimal control. This construction is based on the straightforward expansion of the cdf of $Y$:

$$(13) \qquad F(y) = \sum_{j=1}^m P_j(y)(\alpha_j - \alpha_{j-1}),$$

where $P_j(y)$ is the conditional probability

$$(14) \qquad P_j(y) = \mathbb{P}(Y \leq y | Z \in (z_{\alpha_{j-1}}, z_{\alpha_j}]).$$

The quantiles of $Z$ are known, so the estimation of $F(y)$ is reduced to the estimations of the conditional probabilities. The Poststratified Sampling (PS) estimator of $F(y)$ is

$$\hat{F}_{\text{PS}}(y) = \sum_{j=1}^{m} \hat{P}_j(y)(\alpha_j - \alpha_{j-1}),$$

where

$$\hat{P}_j(y) = \frac{\sum_{i=1}^{n} \mathbf{1}_{Z_i \in (z_{\alpha_{j-1}}, z_{\alpha_j}]} \mathbf{1}_{Y_i \leq y}}{\sum_{i=1}^{n} \mathbf{1}_{Z_i \in (z_{\alpha_{j-1}}, z_{\alpha_j}]}}.$$

The PS estimator can be written as a weighted average of $\mathbf{1}_{Y_j \leq y}$ as well. It can also be interpreted as a CV estimator with $g^j(Z) = \mathbf{1}_{Z \leq z_{\alpha_j}}$, $j = 1, \ldots, m$, as control variates. Its variance is

$$(15) \qquad \text{Var}(\hat{F}_{\text{PS}}(y)) = \frac{1}{n} \sum_{j=1}^{m} (\alpha_j - \alpha_{j-1})[P_j(y) - P_j^2(y)] + O\left(\frac{1}{n^2}\right).$$

Using Gaussian examples, Hesterberg and Nelson (1998) have shown that the optimal variance reduction (the one achieved with $g^*$) can be almost achieved with the discrete approximation with two or three strata. Based on numerical simulations, the authors recommend to choose the cutpoint $\alpha_1 = \alpha$ for the PS strategy with two strata. They also apply the strategy with three strata on some particular examples. In the next section we will show that we can go beyond the variance reduction obtained with the optimal control $g^*(Z)$ or its approximations by using the reduced model in a different way.

**3. Quantile estimation by Controlled Stratification (CS).** The use of a reduced model to estimate directly the quantiles may be not efficient. Indeed, as mentioned in the introduction, the reduced model is usually a metamodel or a response surface that has been calibrated to mimic the response of the complete model $f(X)$ for typical realizations of $X$, and not to predict the response $f(X)$ for exceptional realizations of $X$. This is precisely what is sought when the purpose is to estimate quantiles. Besides, it is very difficult to give an estimate of the error when only the reduced model is used to estimate the quantiles.

In this section we exploit the existence of a reduced model $Z = f_r(X)$ in a different manner than the CV strategy. The idea of the previous section was to use the reduced model as a control variate, or equivalently, poststratification, without modifying the sampling. The idea of this section is to use it in order to implement nonproportional stratified sampling in which we do modify the sampling by rejection. The rough idea is to generate many realizations of $X$, to evaluate the corresponding reduced responses $f_r(X)$, and to accept/reject the realizations depending on the responses $f_r(X)$. The complete model $f$ will be used only with the accepted realizations. We can therefore enforce the numbers of realizations of $X$ such that $f_r(X)$ lie in prescribed intervals, and increase the numbers of realizations in the more important intervals.

3.1. *Estimation of the distribution function.* Let us choose $m+1$ cut-points $0 = \alpha_0 < \alpha_1 < \cdots < \alpha_m = 1$ and denote by $-\infty = z_{\alpha_0} < z_{\alpha_1} < \cdots < z_{\alpha_m} = \infty$ the corresponding quantiles of $Z$. As noted in the previous section, the cdf of $Y$ can be expanded as (13), so the estimation of $F(y)$ is reduced to the estimations of the conditional probabilities $P_j(y)$ defined by (14). Let us fix a sequence of integers $N_1, \ldots, N_m$ such that $\sum_{j=1}^m N_j = n$, where $n$ is the total number of simulations using the complete model $f$. For each $j$, we use the rejection method to sample $N_j$ realizations of the input r.v. $(X_i^{(j)})_{i=1,\ldots,N_j}$, such that the reduced output r.v. $Z_i^{(j)} = f_r(X_i^{(j)})$ lies in the interval $(z_{\alpha_{j-1}}, z_{\alpha_j}]$. For each of these $N_j$ realizations, the output $Y_i^{(j)} = f(X_i^{(j)})$ is computed. The conditional probability $P_j(y)$ can be estimated by

$$\hat{P}_j(y) = \frac{1}{N_j} \sum_{i=1}^{N_j} \mathbf{1}_{Y_i^{(j)} \leq y},$$

which gives the CS estimator of $F(y)$,

$$(16) \qquad \hat{F}_{\mathrm{CS}}(y) = \sum_{j=1}^m \hat{P}_j(y)(\alpha_j - \alpha_{j-1}).$$

The estimator $\hat{F}_{\mathrm{CS}}(y)$ is unbiased and its variance is

$$(17) \qquad \mathrm{Var}(\hat{F}_{\mathrm{CS}}(y)) = \sum_{j=1}^m \frac{(\alpha_j - \alpha_{j-1})^2}{N_j} [P_j(y) - P_j^2(y)].$$

If the number $m$ of strata is fixed, if $(\beta_j)_{j=1,\ldots,m}$ is a sequence of positive real numbers such that $\sum_{j=1}^m \beta_j = 1$, and if we choose $N_j = [n\beta_j]$, where $[x]$ is the integer closest to $x$, then the estimator $\hat{F}_{\mathrm{CS}}(y)$ is asymptotically normal:

$$(18) \qquad \sqrt{n}(\hat{F}_{\mathrm{CS}}(y) - F(y)) \stackrel{n \to \infty}{\longrightarrow} \mathcal{N}(0, \sigma_{\mathrm{CS}}^2),$$

$$\sigma_{\mathrm{CS}}^2 = \sum_{j=1}^m \frac{(\alpha_j - \alpha_{j-1})^2}{\beta_j} [P_j(y) - P_j^2(y)].$$

We first try to estimate the complete cdf $F(y)$ for all $y \in \mathbb{R}$.

When $Z$ is independent of $Y$ (which means that there is no control), then we have $P_j(y) = F(y)$ and

$$\text{Var}(\hat{F}_{\text{CS}}(y)) = \left[\sum_{j=1}^{m} \frac{(\alpha_j - \alpha_{j-1})^2}{N_j}\right] \times [F(y) - F(y)^2]. \tag{19}$$

If we use a proportional allocation in the strata $\beta_j = \alpha_j - \alpha_{j-1}$, then $N_j = [(\alpha_j - \alpha_{j-1})n]$ and we find that the variance of the CS estimator is, modulo the rounding errors, equal to $\frac{1}{n}[F(y) - F(y)^2]$, which is the variance of the empirical estimator.

When the r.v. $Z$ is an increasing function of $Y$ (i.e., to say, $Z$ controls completely $Y$), then we obtain

$$\text{Var}(\hat{F}_{\text{CS}}(y)) = \frac{(\alpha_{j_0} - \alpha_{j_0-1})^2}{N_{j_0}}[p_{j_0}(y) - p_{j_0}(y)^2] \leq \frac{(\alpha_{j_0} - \alpha_{j_0-1})^2}{4 N_{j_0}},$$

where $j_0$ is such that $y \in (y_{\alpha_{j_0-1}}, y_{\alpha_{j_0}}]$. If we choose equiprobable strata $\alpha_j = j/m$ and proportional sampling $\beta_j = 1/m$, then $\text{Var}(\hat{F}_{\text{CS}}(y)) \leq 1/(4mn)$. The variance of the CS estimator has therefore been reduced by a factor of the order of $1/m$.

Let us now look for the estimation of the tail of cdf $F(y)$, in the region where $F(y) \simeq 1 - \delta$ with $0 < \delta \ll 1$. If $Z$ and $Y$ have positive correlation, then it is clear that we should allocate more simulation points in the tail of the reduced model $Z$, so as to increase the number of realizations that are potentially relevant.

As an example, we can choose $m = 4$, $\alpha_1 = 1/2$, $\alpha_2 = 1 - 2\delta$, $\alpha_3 = 1 - \delta$, $N_j = n/4$ for $j = 1, \ldots, 4$. Note that this particular choice allocates $n/2$ points in the tail of the cdf of $Z$, where $z_{1-2\delta} < Z$.

If $Z$ and $Y$ are independent, then equation (19) shows that this strategy involves an increase of the variance of the estimator by a factor 2: $\text{Var}(\hat{F}_{\text{CS}}(y)) \simeq 2\delta/n$ compared to the empirical estimator $\text{Var}(\hat{F}_{\text{EE}}(y)) \simeq \delta/n$.

If $Z$ and $Y$ are so strongly correlated that the probability of the joint event $Z \leq z_{1-2\delta}$ and $Y > y_{1-\delta}$ is negligible, then equation (19) shows that this strategy involves a variance reduction by a factor smaller than $2\delta$: $\text{Var}(\hat{F}_{\text{CS}}(y)) \leq 2\delta^2/n$. This means that the variance reduction can be very substantial in the case where the variables $Y$ and $Z$ are correlated.

3.2. *Quantile estimation.* Here we consider the problem of the estimation of the $\alpha$-quantile of $Y$, with $\alpha$ close to 1. From the previous results, we can propose the estimator of the $\alpha$-quantile of $Y$ given by

$$\hat{Y}_{\text{CS}}(\alpha) = \inf\{y, \hat{F}_{\text{CS}}(y) > \alpha\},$$

where $\hat{F}_{\mathrm{CS}}(y)$ is the CS estimator of the cdf of $Y$. The estimator $\hat{Y}_{\mathrm{CS}}(\alpha)$ is asymptotically normal,

$$\sqrt{n}(\hat{Y}_{\mathrm{CS}}(\alpha) - y_\alpha) \stackrel{n \to \infty}{\longrightarrow} \mathcal{N}(0, \sigma^2_{\mathrm{CS}}),$$

$$\sigma^2_{\mathrm{CS}} = \frac{\sum_{j=1}^{m} \frac{(\alpha_j - \alpha_{j-1})^2}{\beta_j}[P_j(y_\alpha) - P_j^2(y_\alpha)]}{p^2(y_\alpha)}.$$

If $Z$ and $Y$ are positively correlated, then it is profitable to allocate more points in the cdf tail of $Z$, so as to increase the number of potentially relevant realizations.

In the following we carry out numerical simulations on a toy example in which $n = 200$ and $\alpha = 0.95$. We apply the CS method with $m = 4$ strata described here above ($\alpha_1 = 0.5$, $\alpha_2 = 0.9$, $\alpha_3 = 0.95$, $N_j = n/4$ for $j = 1, \ldots, 4$). We underline that this example is very simple and the reduced model could certainly be improved. In particular, a Gaussian process approach would provide a very good approximation with a few tens of simulations (see Section 5). The reduced model for this toy example has in fact been chosen so as to have approximately the same quality (in terms of correlation coefficients $\rho$ and $\rho_I$) as the one we expect in the case of the real example addressed in Section 5. Our first objective here is to validate the CS strategy on this toy example for which we can check the CS estimations in terms of bias and standard deviation. Our second objective is to show that it can give good results with the parameters $n = 200$ and $\alpha = 0.95$ even in the case in which $\rho_I$ is relatively small, which is the context of the real example addressed in Section 5.

*Toy example. 1D function.* Let us consider the following configuration. $X$ is assumed to be a Gaussian r.v. with mean zero and variance one. The functions $f$ and $f_r$ are given by

(20) $\quad f_r(x) = x^2, \qquad f(x) = 0.95 x^2 [1 + 0.5 \cos(10x) + 0.5 \cos(20x)].$

The quantiles of $Z = f_r(X)$ are given by $z_\alpha = [\Phi^{-1}((1+\alpha)/2)]^2$, where $\Phi$ is the cdf of the $\mathcal{N}(0,1)$-distribution. The quantiles of $Y$ are not known analytically, as can be seen in the plot of the pdf of $Y$ in Figure 1(a), obtained by a series of $5\,10^7$ Monte Carlo simulations. By using these Monte Carlo simulations, we have evaluated the theoretical 0.95-quantile of $Y$: $y_{0.95} = 3.66$, and the correlation coefficient between $Y$ and $Z$: $\rho = 0.84$. The efficiency of the CS method is related to the value of the indicator correlation coefficient $\rho_I$, which can be computed from the simulations and (11): $\rho_I = 0.62$. We compare the CS estimator with the empirical estimator and the CV estimator of the $\alpha$-quantile (Figure 1(b)). One can observe that the quantile is poorly predicted by the empirical estimator, slightly better predicted by the CV estimator, while the CS estimator seems more efficient.

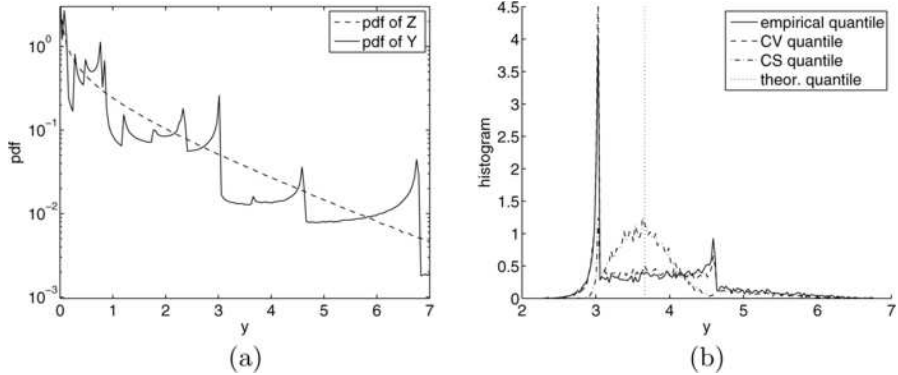

FIG. 1. (a): *Pdf of* $Y = f(X)$ *and* $Z = f_r(X)$ *for* $X \sim \mathcal{N}(0,1)$, *obtained with* $510^7$ *Monte Carlo simulations.* (b): *Estimation of the* $\alpha$-*quantile of the r.v.* $Y = f(X)$ *from a* $n$-*sample, with* $\alpha = 95\%$ *and* $n = 200$. *The three histograms are obtained from three series of* $10^4$ *experiments. The theoretical quantile is* $y_\alpha = 3.66$. *The mean of the* $10^4$ *empirical estimations is* 3.86 *and their standard deviation is* 0.83. *The mean of the* $10^4$ *CV estimations is* 3.74 *and their standard deviation is* 0.744. *The mean of the* $10^4$ *CS estimations is* 3.63 *and their standard deviation is* 0.381.

3.3. *Adaptive controlled stratification (ACS).* We first show that there exists an optimal choice for the allocation of the simulation points in the strata. Let us consider the CS estimation of the cdf of $Y$ as described in Section 3.1. Let us fix $y \in \mathbb{R}$. The CS estimator (16) of $F(y)$ is asymptotically normal and its reduced variance $\sigma^2_{\mathrm{CS}}$ is given by (18). In fact, if the rounding errors are neglected, $\sigma^2_{\mathrm{CS}}/n$ is the variance of the CS estimator $\hat{F}_{\mathrm{CS}}(y)$ for any $n$ by (17).

We note that, if we choose to allocate the simulation points proportionally in the strata, that is, we choose $\beta_j = \alpha_j - \alpha_{j-1}$, then the reduced variance of the CS estimator is equivalent to the reduced variance of the PS estimator (15). The important point is that this proportional allocation is not efficient, as we now show.

If the number $m$ of strata is fixed, as well as the cutpoints $(\alpha_j)_{j=0,\ldots,m}$ and the total number $n$ of simulations, then it is possible to choose the numbers of simulations $[\beta_j n]$ per stratum so as to minimize the variance of the CS estimator. It is well known that an optimal allocation policy for standard stratification exists [Fishman (1996), Glasserman, Heidelberger and Shahabuddin (1998)]. Here we show that an optimal allocation policy exists for CS, in which the strata are determined by the metamodel. By denoting $q_j = (\alpha_j - \alpha_{j-1})^2[P_j(y) - P_j^2(y)]$, the minimization problem

$$\arg\min_\beta \left\{ \sum_{j=1}^m \frac{q_j}{\beta_j} \right\}, \qquad \text{with the constraints } \beta_j \geq 0, \sum_{j=1}^m \beta_j = 1$$

has the unique solution

$$\beta_j^* = \frac{q_j^{1/2}}{\sum_{l=1}^m q_l^{1/2}}, \qquad j = 1, \ldots, m.$$

With this optimal choice the reduced variance of the CS estimator $\hat{F}_{\mathrm{CS}}(y)$ is

(21) $$\sigma_{\mathrm{OCS}}^2 = \left\{ \sum_{j=1}^m (\alpha_j - \alpha_{j-1})[P_j(y) - P_j^2(y)]^{1/2} \right\}^2.$$

Note that the optimal allocation $\beta_j^*$ depends on $y$ in general, which means that it is not possible to propose an allocation that is optimal for the estimation of the whole cdf of $Y$. However, we can observe the following:

(1) If the control is weak, then $P_j(y)$ depends weakly on $y$, and the optimal allocation is then $\beta_j^* = \alpha_j - \alpha_{j-1}$.

(2) If the control is strong, then we should allocate more simulations in the strata $(z_{\alpha_{j-1}}, z_{\alpha_j}]$ around $F(y)$.

For instance, let us assume a very strong control, in the sense that $Z = \psi(Y)$ is an increasing function of $Y$. Then

$$P_j(y) = \mathbb{P}(Y \leq y | Z \in (z_{\alpha_{j-1}}, z_{\alpha_j}]) = \mathbb{P}(Y \leq y | Y \in (\psi^{-1}(z_{\alpha_{j-1}}), \psi^{-1}(z_{\alpha_j})])$$

is equal to 1 if $z_{\alpha_j} \leq \psi(y)$ [i.e., $\alpha_j \leq F(y)$] and to 0 if $z_{\alpha_{j-1}} > \psi(y)$ [i.e., $\alpha_{j-1} > F(y)$]. In these two cases, $q_j$ and the optimal $\beta_j^*$ are zero, and all simulations should be allocated in the stratum $(z_{\alpha_{j_0-1}}, z_{\alpha_{j_0}}]$, for which $\alpha_{j_0-1} < F(y) \leq \alpha_{j_0}$. Of course, the very strong control assumed here is not realistic, but this example clearly illustrates the optimal allocation of simulations in the different strata.

We now know that there exists an optimal allocation of the $n$ simulations in the $m$ strata. This allocation depends on the $P_j(y)$, which are the quantities that we want to estimate. We can therefore propose an adaptive procedure:

(1) First apply the CS method with $\widetilde{n} = n^\gamma$ simulations, $\gamma \in (0, 1)$, and an a priori choice of the $\beta_j$. We then obtain a first estimation of the conditional probabilities $P_j(y)$:

$$\widetilde{P}_j(y) = \frac{1}{[\beta_j \widetilde{n}]} \sum_{i=1}^{[\beta_j \widetilde{n}]} \mathbf{1}_{Y_i^{(j)} \leq y}.$$

(2) Estimate the optimal allocation $\beta_j^*$ by

$$\widetilde{\beta}_j = \frac{(\alpha_j - \alpha_{j-1})[\widetilde{P}_j(y) - \widetilde{P}_j(y)^2]^{1/2}}{\sum_{l=1}^m (\alpha_l - \alpha_{l-1})[\widetilde{P}_l(y) - \widetilde{P}_l(y)^2]^{1/2}}.$$

(3) Carry out the $n - \widetilde{n}$ last simulations by allocating the simulations in each stratum in order to achieve the estimated optimal number $[n\widetilde{\beta}_j]$ for all $j$.

(4) Estimate $P_j(y)$ and $F(y)$ by

$$\hat{P}_j(y) = \frac{1}{[\widetilde{\beta}_j n]} \sum_{i=1}^{[\widetilde{\beta}_j n]} \mathbf{1}_{Y_i^{(j)} \leq y}, \qquad \hat{F}_{\mathrm{ACS}}(y) = \sum_{j=1}^{m} \hat{P}_j(y)(\alpha_j - \alpha_{j-1}).$$

The ACS estimator is unbiased conditionally on $\widetilde{\beta}_j > 0$ for the $j$'s such that $\beta_j^* > 0$. The probability of the complementary event is of the order of $\exp(-c\widetilde{n})$ which can be usually neglected. The ACS estimator $\hat{F}_{\mathrm{ACS}}(y)$ is asymptotically normal,

(22)
$$\sqrt{n}(\hat{F}_{\mathrm{ACS}}(y) - F(y)) \stackrel{n \to \infty}{\longrightarrow} \mathcal{N}(0, \sigma^2_{\mathrm{ACS}}),$$

$$\sigma^2_{\mathrm{ACS}} = \left\{ \sum_{j=1}^{m} (\alpha_j - \alpha_{j-1})[P_j(y) - P_j^2(y)]^{1/2} \right\}^2.$$

The expression of the reduced variance $\sigma^2_{\mathrm{ACS}}$ is the same as (21), which is the one of the CS estimator with the optimal allocation $\beta_j^*$. The difference is that the variance $\sigma^2_{\mathrm{ACS}}/n$ is only reached asymptotically as $n \to \infty$ in the case of the ACS estimator, while the variance is $\sigma^2_{\mathrm{OCS}}/n$ for all $n$ in the case of the CS estimator with the optimal allocation. Note that the convergence of the ACS estimator is ensured whatever the choice of the positive a priori numbers $\beta_j$. In practice, a good a priori choice will speed up the convergence.

We now present an asymptotic analysis of the variance reduction for the CV, PS, CS and ACS methods in the case of $m = 2$ strata.

In the PS method, or in the CS method if we choose the proportional allocation $\beta_j = \alpha_j - \alpha_{j-1}$, the reduced variance in (18) is

(23) $$\sigma^2_{\mathrm{PS}} = F(y)(1 - F(y))[1 - \rho_I^2],$$

where $\rho_I$ is the correlation coefficient between $\mathbf{1}_{Y \leq y}$ and $\mathbf{1}_{Z \leq z_\alpha}$ defined by (8). $\sigma^2_{\mathrm{PS}}$ is also the reduced variance of the CV estimator (7). Hesterberg and Nelson (1998) have already noticed that the PS and CV estimators are equivalent. In the ACS method, the expression of the reduced variance $\sigma^2_{\mathrm{ACS}}$ in (22) is

$$\sigma^2_{\mathrm{ACS}} = F(y)(1 - F(y))K^2,$$

$$K = \alpha \left[ 1 + \rho_I \left( \frac{(1-\alpha)(1-F(y))}{\alpha F(y)} \right)^{1/2} \right]^{1/2} \left[ 1 - \rho_I \left( \frac{(1-\alpha)F(y)}{\alpha(1-F(y))} \right)^{1/2} \right]^{1/2}$$
$$+ (1-\alpha) \left[ 1 + \rho_I \left( \frac{\alpha F(y)}{(1-\alpha)(1-F(y))} \right)^{1/2} \right]^{1/2}$$

$$\times \left[1 - \rho_I \left(\frac{\alpha(1-F(y))}{(1-\alpha)F(y)}\right)^{1/2}\right]^{1/2}.$$

If we assume that the correlation coefficient $\rho_I$ is small, then we get the following expansion with respect to $\rho_I$:

$$(24) \qquad \sigma_{\text{ACS}}^2 = F(y)(1-F(y))\left[1 - \frac{\rho_I^2}{8F(y)(1-F(y))} + O(\rho_I^3)\right].$$

These results show that the CV, PS, CS and ACS methods involve a variance reduction of the same order when the goal is to estimate the cdf around the median $F(y) \sim 1/2$. However, when the goal is to estimate the cdf tail $F(y) \sim 0$ or $1$, the ACS method gives a larger variance reduction. Of course, the CS method with a nearly optimal allocation policy gives the same performance as the ACS method, but the implementation of this method requires some a priori information on the correlation between $Y$ and $Z$ to guess the correct allocation, while the ACS method finds it.

The expressions that we have just derived also give indications for the choice of the cutpoint $\alpha$. Indeed, the variance reduction is all the larger as the correlation coefficient $\rho_I$ is larger. For instance, if we assume that $Z = \psi(Y)$ is an increasing function of $Y$, which models a very strong control, then one finds

$$\rho_I = \begin{cases} \left[\frac{\alpha(1-F(y))}{(1-\alpha)F(y)}\right]^{1/2}, & \text{if } z_\alpha < \psi(y), \\ \left[\frac{(1-\alpha)F(y)}{\alpha(1-F(y))}\right]^{1/2}, & \text{if } z_\alpha \geq \psi(y). \end{cases}$$

As a function of $\alpha$, this function is maximal when $\alpha = F(y)$. This shows that, if the goal is to estimate the cdf of $Y$ around some $y$, then it is interesting to choose $\alpha = F(y)$.

We can also revisit the asymptotic analysis of the variance reduction for the CV, PS, CS and ACS methods in the case of a large number $m$ of strata. In the PS method and in the CS method with the proportional allocation $\beta_j = \alpha_j - \alpha_{j-1}$, the reduced variance is (18). If the conditional probability $\mathbb{P}(Y \leq y|Z)$ has a continuous density $g^*$ with respect to the Lebesgue measure, then (18) is a Riemann sum that has the following limit as $m \to \infty$:

$$\sigma_{\text{PS}}^2 = \mathbb{E}[\mathbb{P}(Y \leq y|Z) - \mathbb{P}(Y \leq y|Z)^2] = F(y) - \mathbb{E}[\mathbb{P}(Y \leq y|Z)^2].$$

This is actually the reduced variance of the optimal CV estimator, when the optimal control function $g^*$ defined by (12) is used. In the ACS method, the expression of the reduced variance $\sigma_{\text{ACS}}^2$ is (22), which has the following limit as $m \to \infty$:

$$\sigma_{\text{ACS}}^2 = \mathbb{E}[(\mathbb{P}(Y \leq y|Z) - \mathbb{P}(Y \leq y|Z)^2)^{1/2}]^2.$$

The Cauchy–Schwarz inequality clearly shows that the variance reduction is larger for the ACS method than for the optimal CV method using the optimal control function $g^*$.

Whatever the value of $m \geq 2$, we can also use the Cauchy–Schwarz inequality to check that the reduced variance for the PS method (or for the CS method with the proportional allocation)

$$\sigma_{\text{PS}}^2 = \sum_{j=1}^m (\alpha_j - \alpha_{j-1})[P_j(y) - P_j^2(y)]$$

is always larger than the reduced variance (22) for the ACS method.

Finally, it is relevant to estimate the additional computational cost of controlled stratification compared to empirical estimation. It is of the order of $(N_r - n)T_X + N_r T_{f_r}$, where $N_r$ is the number of evaluations of $f_r$ and $X$, $T_X$ is the computational time for the generation of a realization of the input r.v. $X$, and $T_{f_r}$ is the computational time for the call of the function $f_r$. For the CS method with the allocation $\beta_j$, we have

$$\mathbb{E}[N_r] = n \sum_{j=1}^m \frac{\beta_j}{\alpha_j - \alpha_{j-1}} \leq \frac{n}{\min_j(\alpha_j - \alpha_{j-1})},$$

where the last inequality holds uniformly in $\beta$. Besides, $N_r$ has fluctuations of the order of $\sqrt{n}$. The same estimate holds true for the ACS method. In the real example we have in mind (in which the computational time for the function $f$ is 20 minutes), this additional cost is negligible.

3.4. *Quantile estimation by adaptive controlled stratification.* In this section we use the ACS strategy to estimate the $\alpha$-quantile of $Y$, with $\alpha$ close to 1. We propose the following procedure:

(1) First apply the CS method with $\widetilde{n} = n^\gamma$ simulations, $\gamma \in (0, 1)$, and with an a priori allocation policy $\beta_j$, so that a first estimate of the conditional probabilities $P_j(y)$ can be obtained:

$$\widetilde{P}_j(y) = \frac{1}{[\beta_j \widetilde{n}]} \sum_{i=1}^{[\beta_j \widetilde{n}]} \mathbf{1}_{Y_i^{(j)} \leq y}.$$

The corresponding estimators of the cdf and the $\alpha$-quantile of $Y$ are

$$\widetilde{F}(y) = \sum_{j=1}^m (\alpha_j - \alpha_{j-1})\widetilde{P}_j(y), \qquad \widetilde{Y}_\alpha = \inf\{y, \widetilde{F}(y) > \alpha\}.$$

(2) Estimate the optimal allocation $\beta_j^*$ for the estimated $\alpha$-quantile $\widetilde{Y}_\alpha$ by

(25) $$\widetilde{\beta}_j = \frac{(\alpha_j - \alpha_{j-1})[\widetilde{P}_j(\widetilde{Y}_\alpha) - \widetilde{P}_j(\widetilde{Y}_\alpha)^2]^{1/2}}{\sum_{l=1}^m (\alpha_l - \alpha_{l-1})[\widetilde{P}_l(\widetilde{Y}_\alpha) - \widetilde{P}_l(\widetilde{Y}_\alpha)^2]^{1/2}}.$$

(3) Carry out the $n - \widetilde{n}$ final simulations by allocating the simulations in each stratum in order to achieve the estimated optimal number $[\widetilde{\beta}_j n]$.

(4) Estimate $P_j(y)$ and $F(y)$ by

$$\hat{P}_j(y) = \frac{1}{[\widetilde{\beta}_j n]} \sum_{i=1}^{[\widetilde{\beta}_j n]} \mathbf{1}_{Y_i^{(j)} \leq y}, \qquad \hat{F}_{\mathrm{ACS}}(y) = \sum_{j=1}^{m} \hat{P}_j(y)(\alpha_j - \alpha_{j-1}).$$

The ACS estimator of the $\alpha$-quantile $y_\alpha$ is

$$\hat{Y}_{\mathrm{ACS}}(\alpha) = \inf\{y, \hat{F}_{\mathrm{ACS}}(y) > \alpha\}.$$

The estimator $\hat{Y}_{\mathrm{ACS}}(\alpha)$ is asymptotically normal,

$$\sqrt{n}(\hat{Y}_{\mathrm{ACS}}(\alpha) - y_\alpha) \stackrel{n\to\infty}{\longrightarrow} \mathcal{N}(0, \sigma_{\mathrm{ACS}}^2),$$

$$\sigma_{\mathrm{ACS}}^2 = \frac{\{\sum_{j=1}^{m}(\alpha_j - \alpha_{j-1})[P_j(y_\alpha) - P_j^2(y_\alpha)]^{1/2}\}^2}{p^2(y_\alpha)}.$$

To summarize, we have found the following expressions of the reduced variance for the different methods:

- for the empirical estimator,

$$\sigma_{\mathrm{EE}}^2 = \frac{\alpha(1-\alpha)}{p^2(y_\alpha)}.$$

- for the PS estimator or for the CV estimator with the proportional allocation $\beta_j = \alpha_j - \alpha_{j-1}$ [see (10)],

$$\sigma_{\mathrm{PS}}^2 = \frac{\alpha(1-\alpha)}{p^2(y_\alpha)} \times (1 - \rho_I^2).$$

- for the ACS estimator with two strata separated by $\alpha$

$$\sigma_{\mathrm{ACS}}^2 = \frac{\alpha(1-\alpha)}{p^2(y_\alpha)} \times \left(1 - \frac{\rho_I^2}{8\alpha(1-\alpha)} + O(\rho_I^3)\right).$$

Here $\rho_I$ is the correlation coefficient between $\mathbf{1}_{Y \leq y_\alpha}$ and $\mathbf{1}_{Z \leq z_\alpha}$ given by (11). This shows that the CV, PS, CS and ACS methods give variance reductions of the same order when the goal is to estimate quantiles close to the median $\alpha \sim 1/2$. However, when the goal is to estimate large quantiles $\alpha \sim 0$ or 1, the ACS method is much more efficient.

3.5. *Simulations.* We now present a series of numerical simulations that illustrate the theoretical results presented in this paper. These examples are simple and they are used to validate the ACS method when $n = 200$, $\alpha = 0.95$ and the reduced model has poor quality (i.e., $\rho_I$ is small). We will address in Section 5 a real example in which these conditions hold.

*Toy example. 1D function.* Let us revisit the toy example based on the 1D function (20) and look for the $\alpha$-quantile of $Y$ with $\alpha = 95\%$. We compare the performances of the empirical estimator (2), the CV estimator (9) with the control variate $Z$, and the ACS estimator. The ACS method is first implemented with two strata $[0, \alpha_1]$ and $(\alpha_1, 1]$ with the cutpoint $\alpha_1 = \alpha$. We use $\tilde{n} = 2n/10$ simulations for the estimation (25) of the optimal allocation, with $n/10$ simulations in each stratum. The results extracted from a series of 10000 simulations are summarized in Table 1. Note that the ACS method allocates 85% of the simulations in the stratum $[0, 0.95]$, and 15% in the stratum $(0.95, 1]$. This shows that we deal with a configuration where the number of simulations in the stratum $(0.95, 1]$ has to be increased compared to the expected value in the case where there is no control, where only 5% of the simulations should be allocated in the stratum $(0.95, 1]$.

We have also implemented the ACS method with 3 strata $[0, \alpha_1]$, $(\alpha_1, \alpha_2]$, $(\alpha_2, 1]$, with cutpoints $\alpha_1 = 0.85$ and $\alpha_2 = 0.95$. We use $3n/10$ simulations for the evaluation (25) of the optimal allocation, with $n/10$ simulations in each of the three strata. The results extracted from a series of 10000 simulations are presented in Table 1. The variance reduction for the ACS method with three strata is very important. The standard deviation of the ACS estimator is 3 times smaller compared to the empirical estimator of the CV estimator. Note that the optimal allocation should attribute a fraction $\beta_1 < 0.1$ to the first stratum, but the number 0.1 cannot be lowered due to the fact that $n/10$ simulations in the first stratum $[0, \alpha_1]$ were already used in the first step of the estimation.

The previous simulations were carried out with the sample size $n = 2000$. In such a case, the ACS method is robust, in the sense that the choice of the number $\tilde{n}$ of simulations devoted to the estimation of the optimal allocation is not critical. When $n$ is smaller, such as $n = 200$, then the choice of $\tilde{n}$ becomes critical: if $\tilde{n}$ is too small, then the estimation of the optimal allocation may fail during the first step of the ACS method; if $\tilde{n}$ is too large, then $n - \tilde{n}$ may be too small and it may be impossible to allocate the estimated optimal number of simulations to each stratum during the second step of the ACS method. We have applied the ACS method with $n = 200$ to the example 1, and it turns out that the ACS method with $\tilde{n} = n/10$ is still efficient. However, we cannot claim that this will be the case for all applications. The results obtained from a series of 10000 simulations are summarized in Table 2. The standard deviations of the estimations of the $\beta_j$'s are much larger than in the case $n = 2000$, but the quality of the estimation of the optimal allocation is just good enough to allow for a significant variance reduction for the quantile estimation. The standard deviation of the ACS estimator is here 2 times smaller compared to the empirical estimator or the CV estimator. If $n = 100$, then the ACS method fails (in the sense that some simulations give $\tilde{\beta}_1 = 0$), and the CS method with an allocation of the simulation points prescribed by the user should be chosen.

TABLE 1
*Estimation of the $\alpha$-quantile with $\alpha = 0.95$, $n = 2000$, $y_\alpha = 3.66$*

| Method | Quantity | Mean | Standard deviation |
|---|---|---|---|
| Empirical estimation | $\hat{Y}_{\text{EE}}(\alpha)$ | 3.66 | 0.33 |
| CV estimation | $\hat{Y}_{\text{CV}}(\alpha)$ | 3.65 | 0.29 |
| ACS method with 2 strata | $\widetilde{\beta}_1$ | 0.86 | 0.02 |
| $[0, 0.95]$, $(0.95, 1]$ | $\hat{Y}_{\text{ACS}}(\alpha)$ | 3.65 | 0.28 |
| ACS method with 3 strata | $\widetilde{\beta}_1$ | 0.10 | 0.02 |
| $[0, 0.85]$, $(0.85, 0.95]$, $(0.95, 1]$ | $\widetilde{\beta}_2$ | 0.58 | 0.02 |
| | $\widetilde{\beta}_3$ | 0.32 | 0.01 |
| | $\hat{Y}_{\text{ACS}}(\alpha)$ | 3.65 | 0.12 |

TABLE 2
*Estimation of the $\alpha$-quantile with $\alpha = 0.95$, $n = 200$, $y_\alpha = 3.66$*

| Method | Quantity | Mean | Standard deviation |
|---|---|---|---|
| Empirical estimation | $\hat{Y}_{\text{EE}}(\alpha)$ | 3.88 | 0.83 |
| CV estimation | $\hat{Y}_{\text{CV}}(\alpha)$ | 3.73 | 0.74 |
| ACS method with 3 strata | $\widetilde{\beta}_1$ | 0.14 | 0.16 |
| $[0, 0.85], (0.85, 0.95], (0.95, 1]$ | $\widetilde{\beta}_2$ | 0.55 | 0.11 |
|  | $\widetilde{\beta}_3$ | 0.31 | 0.06 |
|  | $\hat{Y}_{\text{ACS}}(\alpha)$ | 3.62 | 0.38 |

**4. Quantile estimation with Controlled Importance Sampling (CIS).** We consider the same problem as in the previous sections. In this section we show that the reduced model can be used to help design a biased distribution of the input r.v. $X$ in order to implement an efficient importance sampling (IS) strategy. The standard IS method consists in simulating the $n$-sample of the r.v. $X$ according to a biased distribution, and to multiply the output by a likelihood ratio to recover an unbiased estimator. In the case in which the biased distribution favors the occurrence of the event of importance, the variance of the estimator can be drastically reduced compared to the standard empirical Monte Carlo estimator. Adaptive versions of the IS procedure have been proposed and studied, whose principle is to estimate first a "good" biased distribution that is to say, a distribution that properly favors the occurrence of the event of importance, before using this biased distribution as in the standard IS estimation. We will propose an estimator of the cdf of $Y$ first, then an estimator of the $\alpha$-quantile of $Y$, by a controlled importance sampling (CIS) procedure. In this procedure, the reduced model $f_r$ and the associated r.v. $Z = f_r(X)$ are used to determine the biased distribution, while the complete model $f$ is used to perform the estimation.

4.1. *Estimation of the distribution function.* An IS estimator of the cdf of $Y = f(X)$ is

$$(26) \qquad \hat{F}_{\mathrm{IS}}(y) = \frac{1}{n} \sum_{i=1}^{n} \mathbf{1}_{f(X_i) \leq y} \frac{q_{\mathrm{ori}}(X_i)}{q(X_i)}, \qquad X_i \sim q,$$

where $q_{\mathrm{ori}}$ is the original pdf of $X$ and $q$ is the biased pdf chosen by the user. In practice, it can be useful to use a variant in which the denominator $n$ in (26) is replaced by $\sum_{i=1}^{n} q_{\mathrm{ori}}(X_i)/q(X_i)$, in order to enforce $\hat{F}_{\mathrm{IS}}(y) \to 1$ as $y \to \infty$. Other alternatives can be found in Hesterberg (1995). The estimator $\hat{F}_{\mathrm{IS}}(y)$ converges almost surely to $F(y)$ when $n \to \infty$. In fact, the estimator $\hat{F}_{\mathrm{IS}}(y)$ is unbiased as soon as the support of the original pdf $q_{\mathrm{ori}}$ is included in the support of the biased pdf $q$. The variance of (26) is given by

$$(27) \qquad \mathrm{Var}[\hat{F}_{\mathrm{IS}}(y)] = \frac{1}{n} \left( \int \frac{\mathbf{1}_{f(x) \leq y} q_{\mathrm{ori}}(x)^2}{q(x)} \, dx - F(y)^2 \right).$$

The IS can involve a dramatic variance reduction compared to the standard empirical estimator if the biased pdf $q$ is properly chosen. The variance of $\hat{F}_{\mathrm{IS}}(y)$ is minimal [Rubinstein (1981)] when the biased pdf is taken to be equal to the optimal pdf defined by

$$(28) \qquad q^*(x) = \frac{\mathbf{1}_{f(x) \leq y} q_{\mathrm{ori}}(x)}{\int \mathbf{1}_{f(x') \leq y} q_{\mathrm{ori}}(x') \, dx'}.$$

This result cannot be directly used to perform the simulations because the normalizing constant of the optimal pdf is the quantity that is sought. However, this remark gives the basis of an adaptive procedure where the optimal density is estimated.

The parametric approach for the adaptive IS approach is the following one. We first choose a family of pdfs $\mathcal{Q} = \{q_\gamma; \gamma \in \Gamma\}$ and we then try to estimate the parameter $\gamma$. In this section we assume that the family of pdfs $\mathcal{Q}$ is parameterized by the first two moments: $\gamma = (\lambda, C): \lambda \in \mathbb{R}^d$ is the expectation and $C \in \mathcal{M}_d(\mathbb{R})$ is the covariance matrix of $X$ when the pdf of $X$ is $q_\gamma$.

The strategy to determine the best biased density in the family $\mathcal{Q}$ is based on the following remark. The theoretical optimal density is $q^*$ and it is given by (28). The expectation and covariance matrix of the random vector $X$ under $q^*$ are

$$(29) \quad \lambda^* = \frac{\int x \mathbf{1}_{f(x) \leq y} q_{\mathrm{ori}}(x) \, dx}{\int \mathbf{1}_{f(x) \leq y} q_{\mathrm{ori}}(x) \, dx} \quad \text{and} \quad C^* = \frac{\int x x^t \mathbf{1}_{f(x) \leq y} q_{\mathrm{ori}}(x) \, dx}{\int \mathbf{1}_{f(x) \leq y} q_{\mathrm{ori}}(x) \, dx} - \lambda^* \lambda^{*t}.$$

The idea is to choose in the family $\mathcal{Q}$ the pdf $q_\gamma$ which has expectation $\lambda^*$ and covariance matrix $C^*$, that is to say, we choose the pdf $q_{\gamma^*}$ with $\gamma^* = (\lambda^*, C^*)$.

The problem is now reduced to the estimation of $\lambda^*$ and $C^*$. If we assume that the reduced model is so cheap that we can use as many simulations based on $f_r$ as desired, then we can estimate $\lambda^*$ and $C^*$ by

$$
(30) \quad \begin{cases} \hat{\lambda} = \dfrac{\sum_{i=1}^{\tilde{n}} X_i \mathbf{1}_{Z_i \leq y} q_{\text{ori}}(X_i)/q_0(X_i)}{\sum_{i=1}^{\tilde{n}} \mathbf{1}_{Z_i \leq y} q_{\text{ori}}(X_i)/q_0(X_i)}, \\ \hat{C} = \dfrac{\sum_{i=1}^{\tilde{n}} X_i X_i^t \mathbf{1}_{Z_i \leq y} q_{\text{ori}}(X_i)/q_0(X_i)}{\sum_{i=1}^{\tilde{n}} \mathbf{1}_{Z_i \leq y} q_{\text{ori}}(X_i)/q_0(X_i)} - \hat{\lambda}\hat{\lambda}^t, \end{cases} \quad X_i \sim q_0,
$$

where $q_0$ is an a priori pdf chosen by the user. If no a priori information is available, then the choice $q_0 = q_{\text{ori}}$ is natural. The estimators $\hat{\lambda}$ and $\hat{C}$ are well defined on $\bigcup_{i=1}^{\tilde{n}}\{Z_i \leq y\}$. For completeness, we can set $\hat{\lambda} = 0$ and $\hat{C} = I_d$ on the complementary event, whose probability is of the form $\exp(-c\tilde{n})$. As $\tilde{n} \to \infty$, the estimators $\hat{\lambda}$ and $\hat{C}$ converge almost surely to

$$\lambda_r^* = \frac{\int x \mathbf{1}_{f_r(x) \leq y} q_{\text{ori}}(x)\,dx}{\int \mathbf{1}_{f_r(x) \leq y} q_{\text{ori}}(x)\,dx} \quad \text{and} \quad C_r^* = \frac{\int xx^t \mathbf{1}_{f_r(x) \leq y} q_{\text{ori}}(x)\,dx}{\int \mathbf{1}_{f_r(x) \leq y} q_{\text{ori}}(x)\,dx} - \lambda_r^* \lambda_r^{*t},$$

which are close to $\lambda^*$ and $C^*$ if $f_r$ is a good enough reduced model. The variance of the CIS estimator of $F(y)$ using the biased pdf determined by the metamodel is (27) with $q = q_{\gamma_r^*}$, $\gamma_r^* = (\lambda_r^*, C_r^*)$. $\hat{F}_{\text{CIS}}$ is asymptotically normal,

$$\sqrt{n}(\hat{F}_{\text{CIS}}(y) - F(y)) \stackrel{n \to \infty}{\longrightarrow} \mathcal{N}(0, \sigma^2_{\text{CIS}}),$$

$$\sigma^2_{\text{CIS}} = \int \frac{\mathbf{1}_{f(x) \leq y} q_{\text{ori}}(x)^2}{q_{\gamma_r^*}(x)}\,dx - F(y)^2.$$

Note that the selected biased pdf depends on $y$. Indeed, it is not possible to propose a biased pdf that is efficient for all values of $y$. This is not surprising, since the principle of the IS method is to favor the realizations that probe a specific region of the state space that is important for the target function whose expectation is sought (here, $x \mapsto \mathbf{1}_{f(x) \leq y}$).

4.2. *Quantile estimation.* In this subsection we look for the $\alpha$-quantile of $Y$. The CIS strategy consists in determining a biased pdf that is efficient for the estimation of the expectation

$$(31) \qquad \mathbb{E}[\mathbf{1}_{f_r(X) \leq z_\alpha}] = \int \mathbf{1}_{f_r(x) \leq z_\alpha} q_{\text{ori}}(x)\,dx = \alpha,$$

where $f_r$ is the reduced model and $z_\alpha$ is the $\alpha$-quantile of $Z$, which is assumed to be known. The determination of a biased pdf $q$ that minimizes the IS estimator of the quantity (31) will give a pdf that probes the important regions for the estimation of the $\alpha$-quantile of $Z$, and also of the $\alpha$-quantile of $Y$ if the reduced model is correlated to the complete computer model.

As in the previous subsection, we will look for the biased pdf in a family $\mathcal{Q}$ of pdfs $q_\gamma$ parameterized by the first two moments $\gamma = (\lambda, C)$. By using only the reduced model, we estimate the parameter $\gamma$ with the estimator (30) with $y = z_\alpha$. Next we apply the IS estimator (26) of the cdf of $Y$ by using the complete model and the biased density $q_{\hat{\gamma}}$, $\hat{\gamma} = (\hat{\lambda}, \hat{C})$. Finally, the estimator of the $\alpha$-quantile is $\hat{Y}_{\text{CIS}}(\alpha) = \inf\{y, \hat{F}_{\text{IS}}(y) > \alpha\}$. It is asymptotically normal,

$$\sqrt{n}(\hat{Y}_{\text{CIS}}(\alpha) - y_\alpha) \stackrel{n\to\infty}{\longrightarrow} \mathcal{N}(0, \sigma^2_{\text{CIS}}),$$

$$\sigma^2_{\text{CIS}} = \frac{1}{p^2(y_\alpha)} \left( \int \frac{\mathbf{1}_{f(x) \leq y_\alpha} q_{\text{ori}}(x)^2}{q_{\gamma^*_r}(x)} \, dx - \alpha^2 \right).$$

In the case where the reduced model is not so cheap, adaptive IS strategies can be used with the reduced model to estimate the parameters of the biased density [Oh and Berger (1992)]: roughly speaking, at generation $k$, the parameter $\gamma_k$ is estimated by using a standard IS strategy using the biased pdf $\gamma_{k-1}$ obtained during the computations of the previous generation.

4.3. *Simulations.* Let us consider the case where $X = (X_1, X_2)$ is a random vector with independent Gaussian entries with zero mean and variance one. The functions $f$ and $f_r$ are given by

(32) $$f_r(x) = |x_1|x_1 + x_2,$$

(33)
$$f(x) = 0.95|x_1|x_1[1 + 0.5\cos(10x_1) + 0.5\cos(20x_1)]$$
$$+ 0.7x_2[1 + 0.4\cos(x_2) + 0.3\cos(14x_2)].$$

The pdf of $Y = f(X)$ and $Z = f_r(X)$ are plotted in Figure 2(a). By using Monte Carlo simulations, we have evaluated the correlation coefficient between $Y$ and $Z$: $\rho = 0.90$. From (11), we have also evaluated the indicator correlation coefficient: $\rho_I = 0.64$. The empirical estimator and the CIS estimator of the $\alpha$-quantile of $Y$ are compared in Figure 2(b). The family $\mathcal{Q}$ consists of the set of two-dimensional Gaussian pdfs parameterized by their means and covariance matrices. The comparison is also made with the CV estimator and the CS estimator and it appears that the variance of the CIS estimator is significantly smaller than the one of the other estimators.

CIS is the best strategy in this example. However, CIS (in the present version) is successful only when one unique important region exists in the state space. For instance, in the case of the 1D function treated in the previous sections (where there are two equally important regions far away from each other due to the parity of the function $f$), the CIS strategy fails in the sense that the algorithm to determine the biased pdf does not converge. The use of mixed pdf models should be considered to overcome this limitation and will be the subject of further research.

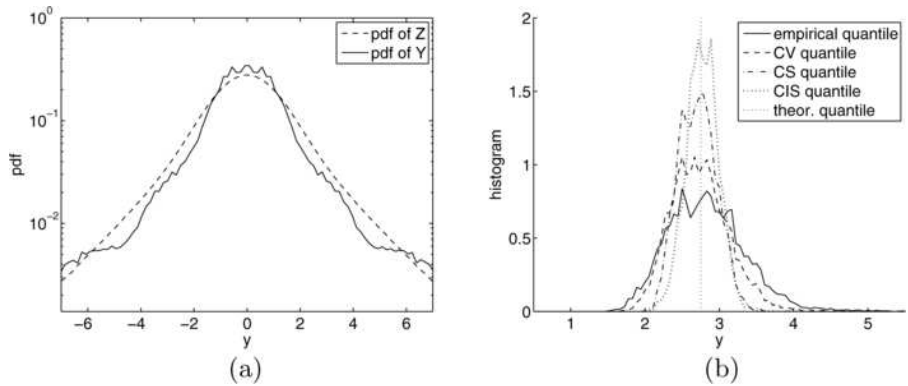

Fig. 2. (a): Pdf of $Y = f(X_1, X_2)$ and $Z = f_r(X_1, X_2)$ for $X_1, X_2 \sim \mathcal{N}(0,1)$. (b): Estimation of the $\alpha$-quantile of $Y$ from a $n$-sample, with $\alpha = 0.95$ and $n = 200$. The four histograms are obtained from four series of $5000$ experiments. The mean of the empirical estimations is $2.83$ and their standard deviation is $0.52$. The mean of the CV estimations is $2.74$ and their standard deviation is $0.38$. The mean of the CS estimations is $2.71$ and their standard deviation is $0.25$. The mean of the CIS estimations is $2.77$ and their standard deviation is $0.21$. The theoretical quantile (obtained from a series of $5\,10^7$ simulations) is $y_\alpha \simeq 2.75$.

**5. Application to a nuclear safety problem.** In this section we apply the controlled stratification and controlled importance sampling methodologies on a complex computer model used for nuclear reactor safety. It simulates a hypothetic thermal-hydraulic scenario: a large-break loss of coolant accident for which the quantity of interest is the peak cladding temperature. This scenario is part of the Benchmark for Uncertainty Analysis in Best-Estimate Modeling for Design, Operation and Safety Analysis of Light Water Reactors [Petruzzi et al. (2004)] proposed by the Nuclear Energy Agency of the Organisation for Economic Co-operation and Development (OCDE/NEA). It has been implemented on the computer code Cathare of the Commissariat à l'Energie Atomique (CEA). In this exercise the 0.95-quantile of the peak cladding temperature has to be estimated with less than 250 computations of the computer model. The CPU time is twenty minutes for each simulation. The complexity of the computer model lies in the high-dimensional input space: 53 random input parameters (physical laws essentially, but also initial conditions, material properties and geometrical modeling) are considered, with normal and log-normal distributions. This number is rather large for the metamodel construction problem.

*Screening and linear regression strategy.*   To simplify the problem, we apply first a screening technique, based on a supersaturated design [Lin (1993)] with 30 numerical experiments. This leads to the determination of the five most influential input parameters ($UO_2$ conductivity $X_{19}$, film boiling heat transfer coefficient $X_{44}$, axial peaking factor $X_9$, critical heat flux $X_{42}$ and $UO_2$ specific heat $X_{20}$). Then a stepwise regression procedure has been applied on the 30 experiments to obtain five additional input parameters and a linear regression procedure allows us to obtain a coarse linear metamodel of degree one:

$$f_r(X) = 660.3 - 61.79X_2 + 6.141X_6 + 589.9X_9 + 80.82X_{11} - 404.5X_{19}$$
$$+ 264.2X_{20} - 27.06X_{35} + 6.161X_{37} - 255.7X_{42} - 31.99X_{44}.$$

Note that the present strategy for the metamodel construction is relatively basic and not devoted to maximize $\rho_I$. Other strategies based on $L^1$ penalization techniques such as Lasso (least absolute shrinkage and selection operator) could be considered to fit the regression model [Tibshirani (1996), Hastie, Tibshirani and Friedman (2001)].

*Controlled stratification.*   A first test with controlled stratification with 200 simulations was performed, which gave the following quantile estimation: $928°C$ with a bootstrap-estimated standard deviation of $7°C$, while the quantile estimation from the metamodel is $932°C$. A second test with controlled stratification with 200 simulations was performed, which gave the following quantile estimation: $929°C$ with a bootstrap-estimated standard deviation of $10°C$.

*Controlled importance sampling.*   A biased distribution for the 10 important parameters of the metamodel has been obtained as follows. The original distributions of these independent parameters are normal or log-normal. We have considered a parametric family of biased pdfs with the same forms as the original ones, and we have selected their means and variances by (30) with $y = z_\alpha$ and $q_0$ equals to the original pdf. A first test with controlled importance sampling with 200 simulations was performed, which gave the following quantile estimation: $929°C$ with a bootstrap-estimated standard deviation of $10°C$, while the quantile estimation from the metamodel is $932°C$. A second test with controlled importance sampling with 200 simulations was performed, which gave the following quantile estimation: $924°C$ with a bootstrap-estimated standard deviation of $8°C$.

*Empirical estimation.*   A test sample of 1000 additional computations (with input parameters chosen randomly) was then carried out. We have first used this random sample to check the quality of the metamodel. We have found that $\rho = 0.66$, $R^2 = 0.09$ and $\rho_I = 0.54$, which shows that the metamodel has poor quality (as it could have been expected). We also used the random sample to get empirical estimations of the quantile. For the full sample $n = 1000$ the empirical quantile estimation is $928°$C with a standard deviation of $6°$C. For $n = 200$ the empirical quantile estimation is $926°$C with a standard deviation of $12°$C. It thus appears that the controlled stratification estimator and the controlled importance sampling estimator performed with 230 simulations (30 for the screening and 200 for the controlled estimation) have better performances than the empirical estimator with 200 simulations, and have performances close to the empirical estimator with 1000 simulations. This shows that controlled stratification and controlled importance sampling can be used to substantially reduce the variance of quantile estimation, in the case in which a small number of simulations is allowed but a reduced cheap model is available, even if this reduced model has poor quality.

*Gaussian process (Gp) strategies.* In order to compare our approach (crude initial screening step before controlled stratification step) to a strategy including a more involved metamodel construction step, we propose to show some results obtained with a Gp approach [Sacks et al. (1989), Schonlau and Welch (2005)].

- First, we perform a numerical experimental design of 200 Cathare code simulations. We choose a maximin Latin hypercube sampling design, well adapted to the Gp model construction [Fang, Li and Sudjianto (2006)]. This difficult fit (due to the high dimensionality and small database) can be realized thanks to the algorithm of Marrel et al. (2008), specifically devoted to this situation. The obtained Gp model contains a linear regression part (including 15 input variables) and a generalized exponential covariance part (including 7 input variables). We use the test sample of 1000 additional computations to check the predictor quality of this new metamodel: $\rho = 0.84$, $R^2 = 0.70$ and $\rho_I = 0.73$. As expected, the quality of this Gp model is much higher than the crude one. However, a brute-force Monte Carlo estimation (with $10^6$ computations) of the 0.95-quantile using the predictor of this metamodel gives $917°$C, which underestimates the "true" quantile ($928°$C). A better strategy, which could be applied in a future work, would be to choose sequentially the specific design points to improve the Gp fit around the quantile, as in Oakley (2004).
- As a second comparison, we propose to perform the controlled stratification process with the predictor of a Gp model. We fit a Gp model with a smaller number of runs than the previous one, keeping other runs for the controlled stratification step. We choose a maximin Latin hypercube sampling design with 100 design points. Below this sampling size, Gp fitting becomes unfeasible because of the large dimensionality of our problem (53 input variables). The obtained Gp model contains a linear regression part (including 7 input variables) and a generalized exponential covariance part (including 6 input variables). The quality of this Gp model is measured via the test sample and gives $\rho = 0.82$, $R^2 = 0.66$ and $\rho_I = 0.37$. The Gp predictivity is rather good but, compared to the previous one, the $\rho_I$ value shows a strong deterioration around the 0.95-quantile (the Gp model 0.95-quantile is 912°C). Using the predictor of this Gp model, the controlled stratification with 200 simulations gives the following quantile estimation: 917°C with a standard deviation of 9°C. This relatively poor and biased result confirms the importance of $\rho_I$ in the controlled stratification process: quantile estimation with a coarse metamodel (linear model of degree one with $R^2 = 0.09$), but adequate near the quantile region, gives better results than quantile estimation with a refined metamodel (Gp model with $R^2 = 0.66$), but inadequate near the quantile region.

TABLE 3
*Estimation of the 0.95-quantile for the nuclear safety problem. Gp(100) [resp., Gp(200)] is the Gp model estimated from 100 (resp., 200) design points. Mm(30) is the metamodel $f_r$ estimated from 30 numerical experiments*

| Method | Quantile estimation | Standard deviation estimated by bootstrap |
|---|---|---|
| EE from code ($n=1000$) | **928** | 6 |
| EE from code ($n=200$) | 926 | 12 |
| EE from Mm(30) ($n=10^6$) | 932 | $\sim 0$ |
| EE from Gp(100) ($n=10^6$) | 912 | $\sim 0$ |
| EE from Gp(200) ($n=10^6$) | 917 | $\sim 0$ |
| CS with Mm(30) test 1 ($n=200$) | 928 | 7 |
| CS with Mm(30) test 2 ($n=200$) | 929 | 10 |
| CS with Gp(100) ($n=200$) | 917 | 9 |
| CIS test 1 ($n=200$) | 929 | 10 |
| CIS test 2 ($n=200$) | 924 | 8 |

Table 3 summarizes all the results we have shown in this section. Other experiments, that will be shown in a future paper, have been made to compare different choices about the strata (number and locations).

**6. Conclusion.** In this paper we have proposed and discussed variance reduction techniques for estimating the $\alpha$-quantile of a real-valued r.v. $Y$ in the case in which:

- the r.v. $Y = f(X)$ is the output of a CPU time expensive computer code with random input $X$,

- the auxiliary r.v. $Z = f_r(X)$ can be used at essentially free cost, where $f_r$ is a metamodel that is a coarse approximation of $f$.

Our goal was to exploit the metamodel to obtain better control variates, stratification or importance sampling than could be obtained without it.

First, we have presented already known variance reduction techniques based on the use of $Z$ as a control variate (CV). The CV methods allow a variance reduction of the quantile estimator by associating to each of the $n$ simulations $Y_i = f(X_i)$ weights that depend on $Z_i = f_r(X_i)$. In the CV methods, $n$ realizations $(X_i)_{i=1,\ldots,n}$ are generated and the corresponding $n$ outputs $f(X_i)$ and $f_r(X_i)$ are computed.

Second, we have developed an original controlled stratification (CS) method, that consists in accepting/rejecting the realizations of the input $X$ based on the values of $f_r(X)$. A large number of realizations of the input $X$ and a large number of evaluations of $f_r$ are used, compared to the CV methods, but the number of evaluations of the complete model $f$ is fixed. In the adaptive controlled stratification (ACS) method, the realizations of the random input $X_i$ are sampled in strata determined by the reduced model $f_r$, and the number of simulations allocated to each stratum is optimized dynamically. The variance reduction can be very substantial. By a theoretical analysis of the asymptotic variance of the estimator and by numerical simulations, we have found that, if $n$ is large enough, the ACS method is the most efficient one. Note that the a priori choice of the parameters for the CS and ACS strategies (choice of $\tilde{n}$, $m$ and $\beta_j$) plays no role in the asymptotic regime $n \to \infty$. However, for $n = 200$, for instance, it plays a primary role. In this paper a toy example with a metamodel that has the same quality (in terms of correlation coefficients) as the one we have in the real example has been used to validate the parameters of the CS strategy. For the time being, we have the feeling that it is the only reasonable strategy when $n$ is not large enough to apply the asymptotic results.

Third, a controlled importance sampling (CIS) strategy has been analyzed, where the biased pdf for the CIS estimator is estimated by intensive simulations using only the reduced model. The variance reduction can be significant. However, an important condition in the present version is that only one important region exists in the state space. The use of mixed pdf models should be considered to overcome this limitation.

The methods presented in this paper suppose the availability of a reduced model or a metamodel. If it is not available, then the construction of a metamodel using linear regression strategies or Gaussian process strategies or $L^1$-penalization strategies is necessary. However, it seems to be sufficient to have a crude approximation of the computer model. In industrial practice, it is often the case due to the nonlinear effects, the high dimensionality of the inputs and the limited numbers of computer experiments [Fang, Li and Sudjianto (2006), Volkova, Iooss and Van Dorpe (2008)]. We can note also that a great advantage of these methods is that it is very easy to carry out the simulations on a parallel computer, with as many nodes as calls of the complex code $f$. One possible investigative way to improve our quantile estimation strategies for the applications would be to optimize the number of runs devoted to the metamodel construction and the number of runs devoted to the quantile estimation. Furthermore, the computer runs of the first step of the ACS method can also serve to update the metamodel; then this refined metamodel can be used in the second step. A further improvement would be to update the metamodel $f_r$ as one obtains more values of $f$ (at least occasionally) during the second step. However, this strategy goes against the parallelization of the method and one should be cautious and conservative in order to avoid bias, but it is certainly an interesting direction of research.

The different tests performed on our industrial application have shown that the metamodel quality has to be sufficient near the quantile region. The quality criterion $\rho_I$ has been identified as a good measure of the potential performance of the controlled stratification process. Another quantile estimation technique, the sequential construction of a Gaussian process model [Oakley (2004), Ranjan, Bingham and Michailidis (2008)], is devoted to optimizing the metamodel construction near the quantile region. As a perspective of our work, we will try to apply this technique to our high-dimensional application.

**Acknowledgments.** Part of this work was carried out during the Summer Mathematical Research Center on Scientific Computing (whose French acronym is CEMRACS), which took place in Marseille in July and August 2006 and which was partly supported by the "Simulation Program" (CEA/Nuclear Energy Division). We thank P. Bazin and A. de Crecy who have proposed the Cathare application and R. Phan Tan Luu for useful discussions. We are also particularly grateful to A. Marrel who performed the Gaussian process modeling and to *filao1* who delivered the final computational power.

C. Cannamela  
CEA Cadarache  
DEN/DEC/SESC/LSC  
13108 Saint Paul lez Durance  
France  
and  
SUPELEC  
Département Signaux  
  and Systèmes Electroniques  
Plateau de Moulon  
3 rue Joliot–Curie  
91192 Gif sur Yvette Cedex  
France  
E-mail: claire.cannamela@supelec.fr

J. Garnier  
Université Paris VII  
Laboratoire de Probabilités  
et Modèles Aléatoires  
  and Laboratoire Jacques–Louis Lions  
2 Place Jussieu  
75251 Paris Cedex 05  
France  
E-mail: garnier@math.jussieu.fr

B. Iooss  
CEA Cadarache  
DEN/DER/SESI/LCFR  
13108 Saint Paul lez Durance  
France  
E-mail: bertrand.iooss@cea.fr